\documentclass[12pt]{article}
\usepackage{amsfonts,amssymb,amsmath}
\usepackage[dvips]{epsfig}
\newcounter{tempeq}
\begin{document}
\title{\bf Exponential Generating Functions for the Associated Bessel Functions}
\author{H. Fakhri\thanks{Email: hfakhri@tabrizu.ac.ir}
\\
{\small {\em Department of Theoretical Physics and
Astrophysics, Physics Faculty,}}\\
{\small {\em University of Tabriz, P O Box 51666-16471, Tabriz, Iran}}\\ \\
B. Mojaveri\thanks{Email: bmojaveri@azaruniv.ac.ir}\,\, and \,\, M.A. Gomshi Nobary\thanks{Email: mnobary@razi.ac.ir}\\
{\small {\em  Department of Physics, Faculty of Science, Razi University, }}
\\{\small {\em Kermanshah, Iran }}}
\maketitle
\begin{abstract}
\noindent Similar to the associated Legendre functions,
the differential equation for the associated Bessel functions $B_{l,m}(x)$
is introduced so that its form remains invariant under the transformation
$l\rightarrow -l-1$. A Rodrigues formula for the associated Bessel functions
as squared integrable solutions in both regions $l<0$ and $l\geq 0$ is presented.
The functions with the same $m$ but with different positive and negative values of $l$
are not independent of each other, while the functions with the same $l+m$ ($l-m$)
but with different values of $l$ and $m$ are independent of each other.
So, all the functions $B_{l,m}(x)$ may be taken into account as the union of the
increasing (decreasing) infinite sequences with respect to $l$.
It is shown that two new different types of exponential generating functions are attributed to
the associated Bessel functions corresponding to these rearranged sequences.
\end{abstract}

\section{Introduction and Motivation}
The generating functions have found many applications in the
physical, chemical and mathematical systems. The study of the
important chance process called the branching process
\cite{Harris,Athreya,Apanasovich}, random graphs and complex
networks \cite{Newman}, polymerization kinetics \cite{McLaughlin},
counting problems in combinatorics \cite{Marcus}, are some
applications of the theory of generating functions. The generating
functions are used to obtain expected values (averages), variances,
moments and cumulants of distributions, and also, to establish
relationships between distributions \cite{Wilf}.

The exponential generating functions and their numerous generalizations  have
been alternatively introduced and studied by various methods for the
orthogonal polynomials and special functions (see Refs.
\cite{Kyriakopoulos,Birtwistle1,Birtwistle2,Bacry,Srivastava1,Wang,Srivastava2,Nieto,Messina,
Jeugt,Gangopadhyay,Pathan,Meijer,Durand1,Durand2}). The generating functions as
continuous functions generally describe the convergence of an
infinite summation of given infinite sequences of functions. In this
sense, the special polynomials and functions of a given sequence in
one variable $x$ can be defined as the coefficients in the expansion
of their generating functions. The generating functions
include various useful properties and all
information that is needed to generate the solutions
corresponding to a differential equation or a set of recursion
relations between those solutions. Therefore, generating functions
are very useful to analyze problems involving summations on the infinite sequences of functions
such as coherent states. The application of generating functions for known
orthonormal special functions, allows one to derive a compact formula for the
coherent states. Generating function corresponding to a
given set of special functions is not unique. This manuscript has
been devoted to introducing new generating functions for the
solutions of the differential equation of associated Bessel
functions which can be applied to obtain bound states of some
solvable models in the framework of supersymmetric quantum
mechanics, such as radial bound states of the hydrogen-like atoms
\cite{Fakhri1}. Then, let us remember that Krall and Frink have
first studied the Bessel polynomials in the formalism of
hypergeometric functions \cite{Krall}. Also, some authors have
introduced some generating functions for Bessel polynomials
\cite{Burchnall,Rainville,Brafman}. Moreover, the generating
functions associated with the group-theoretic techniques and the
Stirling numbers of the second kind have been derived for the
special sequences of the generalized Bessel polynomials
\cite{Srivastava3,Srivastava4,Lin}. Two different types of $q$-analogues
of the generating functions for generalized Bessel polynomials have
been calculated in Ref. \cite{Srivastava5} too.

In comparison with the Bessel and Romanovski differential equations \cite{Raposo},
the Hermite, Laguerre and Jacobi ones  from the viewpoint of their
polynomial solutions application to the physics problems, have
attracted much attention until now. However, the Bessel and
Romanovski polynomials have also been applied to obtain the
wavefunctions of some of the physical potentials.
For example we mention the
factorization methods for the differential equation of associated
Bessel polynomials which are applied to obtain the supersymmetric
structures corresponding to the radial bound states of the hydrogen
atom \cite{Fakhri1}. The solutions of the Schr\"odinger equation for
some noncenteral potentials such as hyperbolic Scarf and
trigonometric Rosen-Morse \cite{Raposo,Dutt,Alvarez}, and of the
Klein-Gordon equation with scalar and vector potentials, are
obtained in terms of the Romanovski polynomials \cite{Chao}. The
trigonometric Rosen-Morse potential is an appropriate candidate  to
describe the quark-gluon dynamics in QCD, since, it can be considered
as an appropriate approximation of the combination of the Coulomb, the infinite wall
and the linear potentials. Therefore, the Bessel and Romanovski
nonclassical polynomials have the merit of taking  more into account
in order to derive new relations.

The manuscript has 2 sections. In Section 2, we
introduce the differential equation of associated Bessel functions
$B^{(q,\beta)}_{l,m}(x)$ in terms of the indices $l$ and $m$, in
similarity with associated Legendre functions $P_{l,m}(x)$:
\begin{eqnarray}
\nonumber
(1-x^2)P^{\prime\prime}_{l,m}(x)
-2x\, P^{\prime}_{l,m}(x)
+\left[l(l+1)-\frac{m^2}{1-x^2}\right]
P_{l,m}(x)=0.
\end{eqnarray}
Despite of the $P_{l,m}(x)$'s, the differential equation for the associated Bessel functions is
altered when $m$ is replaced by $-m$. In
conclusion, the function $B^{(q,\beta)}_{l,-m}(x)$ is not another
solution for it. However, the differential equation is invariant
under transformation $l\rightarrow -l-1$, and it, in turn, leads to
considering the solutions as $B^{(q,\beta)}_{l,m}(x)$ with $l<0$, too.
Consequently, we can offer a Rodrigues formula for
$B^{(q,\beta)}_{l,m}(x)$ as the squared integrable solutions with  $l<0$
and $l\geq 0$. Furthermore, simultaneous realization of laddering
equations with respect to $l$ and $m$ by the given Rodrigues
formula in both regions, is considered. In Section 3, it is shown
that the independent solutions $B^{(q,\beta)}_{l,m}(x)$ are
classified in three different types of infinite sequences. The first type
sequences of the associated Bessel functions have the same $l$ but
different $m$. The generating functions corresponding to them can be
followed via the known generating functions of the Bessel polynomials.
The second and the third types of the sequences are constituted by the
independent associated Bessel functions with the same $l+m$ and
$l-m$, respectively. These two later types provide the possibility to view
the set of the independent associated Bessel functions in
two new perspectives, different from the first one. Finally, we calculate two new kinds of
generating functions for any of these types of sequences depending
on whether $l+m$ and $l-m$ are odd or even.

\section{Associated Bessel functions}
This section includes, in addition to review some results of Ref. \cite{Fakhri2},
the formulation of Rodrigues representation for the associated Bessel functions $B^{(q,\beta)}_{l,m}(x)$
with $l<0$.
Let us first remind that the generalized Bessel polynomials of degree $n$ \cite{Nikiforov}, i.e.
\begin{eqnarray}
&&\hspace{-5mm}B^{(\alpha,\beta)}_{n}(x)=
\frac{a_{n}(\alpha,\beta)}{x^{\alpha}e^{-\frac{\beta}{x}}}\left(\frac{d}{dx}\right)^{n}\left(x^{\alpha+2n}e^{\frac{-\beta}{x}}\right)
=a_{n}(\alpha,\beta)\beta^n y_n(x;\alpha+2,\beta),
\end{eqnarray}
are eigenfunctions of the following linear second order differential operator:
\begin{eqnarray}
&&\hspace{-38mm}
x^{-\alpha}e^{\frac{\beta}{x}}\frac{d}{dx}\left(x^{\alpha+2}e^{\frac{-\beta}{x}}
\frac{d}{dx}\right)B^{(\alpha,\beta)}_{n}(x)=n(n+\alpha+1)B^{(\alpha,\beta)}_{n}(x),
\end{eqnarray}
where $a_{n}(\alpha,\beta)$'s are the normalization coefficients.
It must be pointed out that the $y_n(x;\alpha,\beta)$ representation of the generalized Bessel polynomials is
given by \cite{Srivastava4,Lin}
\begin{eqnarray}
&&\hspace{-5mm}y_n(x;\alpha,\beta)=\sum_{k=0}^{n}
\left(\begin{array}{c}n\\k\end{array}\right)
\left(\begin{array}{c}\alpha+n+k-2\\k\end{array}\right)k!\left(\frac{x}{\beta}\right)^k=
\left(\frac{-x}{\beta}\right)^{n}n!L_{n}^{(1-\alpha-2n)}\left(\frac{\beta}{x}\right).
\end{eqnarray}
For $\beta>0$ and $\alpha<-2$, the generalized Bessel polynomials are
orthoghonal and square integrable  with respect to the weight
function $x^{\alpha}e^{\frac{-\beta}{x}}$ in the interval
$0\leq x<\infty$. Choosing $n=l-m+\frac{q}{2}$ and
$\alpha=2m-q$ in the differential equation (2), it is
straightforward to show that the associated Bessel functions \cite{Fakhri2}
\begin{eqnarray}
&&\hspace{-16mm}B^{(q,\beta)}_{l,m}(x):=\frac{a_{l,m}(q,\beta)}{a_{l-m+\frac{q}{2}}(2m-q,\beta)}x^{m}
B^{(2m-q,\beta)}_{l-m+\frac{q}{2}}(x)=\frac{a_{l,m}(q,\beta)}{x^{m-q}e^{\frac{-\beta}{x}}}
\left(\frac{d}{dx}\right)^{l-m+\frac{q}{2}}\left(x^{2l}e^{\frac{-\beta}{x}}\right),
\end{eqnarray}
with $l-m+\frac{q}{2}\geq 0$ as a non-negative integer, satisfy the following differential equation
\begin{eqnarray}
&&\hspace{-27mm}x^{2}B^{\prime\prime(q,\beta)}_{l,m}+\left[(2-q)x+\beta\right]B^{\prime(q,\beta)}_{l,m}
-\left[\left(l+\frac{q}{2}\right)\left(l-\frac{q}{2}+1\right)+\frac{m\beta}{x}\right]B^{(q,\beta)}_{l,m}=0.
\end{eqnarray}
Similar to the associated Legendre differential equation, since
equation (5) is unaltered when $l$ is replaced by $-l-1$, the
functions $B^{(q,\beta)}_{l,m}(x)$ with negative $l$ are also
another solutions for it. But contrary to the associated Legendre
differential equation, which is quadratic in terms of $m$, the
associated Bessel differential equation (5) is linear in terms of
it. It will be clear from our discussions that the functions
$B^{(q,\beta)}_{l,m}$ with $m\leq l+\frac{q}{2}$ and $m\geq
-l+\frac{q}{2}$ are not normalized by the weight function
$x^{-q}e^{\frac{-\beta}{x}}$. Therefore, we limit our study to
$m-\frac{q}{2}\leq l\leq \frac{q}{2}-m-1$ with $m\leq
\frac{q-1}{2}$, in which $q$ is an integer number. In Fig. 1, we
have schematically shown the 2-fold hierarchy of all the associated
Bessel functions corresponding to $q=6$ as the points $(l, m)$ on a
flat plane whose horizontal and vertical axes are labeled with $l$
and $m$, respectively. Note that for an odd integer $q$, one of the two
parameters $l$ and $m$ has to be half-integer, and the other has to
be integer. The Rodrigues formula (4) shows that
the associated Bessel functions $B^{(q,\beta)}_{l,m}$ are finite summations
of (not necessarily positive) integer and half-integer powers
of $x$ when $q$ is even and odd, respectively.

Now, in order to formulate the laddering relations with respect to $l$ and $m$, and also to realize the
square integrability condition, it is necessary that we obtain
the highest powers of $x$ in  the associated Bessel functions $B^{(q,\beta)}_{l,m}(x)$
for $l<0$ and $l\geq 0$, respectively:
\begin{eqnarray}
\nonumber
\hspace{0.5mm}
B^{(q,\beta)}_{l,m}(x)=\left\{
\begin{array}{llllrrr}
a_{l,m}(q,\beta)
(-1)^{m-l-\frac{q}{2}}\frac{\Gamma\left(-l-m+\frac{q}{2}\right)}{\Gamma\left(-2l\right)}
x^{l+\frac{q}{2}}+O(x^{l+\frac{q}{2}-1})\hspace{20mm} l<0 &&&&\hspace{-8mm}\mbox{(6a)} \\
a_{l,m}(q,\beta)
(-1)^{-m-l+\frac{q}{2}-1}\beta^{2l+1}\frac{\Gamma\left(l-m+\frac{q}{2}+1\right)}{\Gamma\left(2l+2\right)}
x^{-l+\frac{q}{2}-1}+O(x^{-l+\frac{q}{2}-2})\hspace{4mm}  l\geq 0.&&&&\hspace{-8mm}\mbox{(6b)}
\end{array}\right.
\end{eqnarray}
Note that (6a) is directly derived using (4), while (6b) is followed  from the following relation
and also (6a),
\setcounter{equation}{6}
\begin{eqnarray}
\left(\frac{d}{dx}\right)^{2l+1}\left(x^{2l}e^{-\frac{\beta}{x}}\right)=\beta^{2l+1}x^{-2l-2}e^{-\frac{\beta}{x}}\hspace{28mm}l\geq0.
\end{eqnarray}
The inductive reasoning can be applied to prove the relation (7).
The orthoghonality of the associated Bessel functions for a given
$m$, and also their square integrability for $l<0$ and $l\geq 0$ are
obtained, respectively, as
\begin{eqnarray}
\hspace{-1mm}\int^{\infty}_{0}B^{(q,\beta)}_{l,m}B^{(q,\beta)}_{l^{\prime},m}x^{-q}e^{\frac{-\beta}{x}}dx=\delta_{l\,l^{\prime}}\,a^{2}_{l,m}(q,\beta)
\frac{\Gamma(-l-m+\frac{q}{2})\Gamma(l-m+\frac{q}{2}+1)}{\beta^{-2l-1}\,(\mp2l\mp1)}.
\end{eqnarray}
They follow from the relations (4), (6) and  applying integration by
parts $l-m+\frac{q}{2}$ and $-l-m+\frac{q}{2}-1$ times,
respectively. In the $k$-th stage of these processes, the total
differential expressions become zero, because
$l^{\prime}+m-\frac{q}{2}+k+1<0$ and
$-l^{\prime}+m-\frac{q}{2}+k<0$.

Similar to Ref. \cite{Fakhri2}, the associated Bessel differential equation (5)
can be  simultaneously factorized by the ladder operators
\begin{eqnarray}
&&\hspace{-14mm}A^{\pm}_{l,\,m}=\pm x^{2}\frac{d}{dx}+\left(l\mp\frac{q}{2}\right)x
\pm\frac{\left(l\pm m\mp\frac{q}{2}\right)\beta}{2l}, \nonumber\\
&&\hspace{-14mm}
A^{\pm}_{m}=\pm x\frac{d}{dx}-\frac{\beta}{2x}\pm\frac{\beta}{2x}-m+\frac{1}{2}(1+q)\pm\frac{1}{2}(1-q),
\end{eqnarray}
with the following eigenvalues
\begin{eqnarray}
&&\hspace{-14mm}E_{l,m}=\frac{\left(l-m+\frac{q}{2}\right)\left(-l-m+\frac{q}{2}\right)\beta^{2}}{4l^{2}},
\hspace{14mm}{\cal E}_{l,m}=\left(\frac{q}{2}-l-m\right)\left(l-m+\frac{q}{2}+1\right),
\end{eqnarray}
as shape invariance symmetry
equations for the indices ($l$,$m$) and ($l-1$,$m$) as well as ($l$,$m$) and ($l$,$m-1$), respectively:
\setcounter{tempeq}{\value{equation}}
\renewcommand\theequation{\arabic{tempeq}\alph{equation}}
\setcounter{equation}{0} \addtocounter{tempeq}{1}
\begin{eqnarray}
&&\hspace{-10mm}A^{+}_{l,m}A^{-}_{l,m}B^{(q,\,\beta)}_{l,m}(x)=E_{l,m}\,B^{(q,\,\beta)}_{l,m}(x)
\hspace{15mm}A^{-}_{l,m}A^{+}_{l,m}B^{(q,\,\beta)}_{l-1,m}(x)=E_{l,m}\,B^{(q,\,\beta)}_{l-1,m}(x),\\
&&\hspace{-10mm}A^{+}_{m}A^{-}_{m}B^{(q,\,\beta)}_{l,m}(x)={\cal E}_{l,m}\,B^{(q,\beta)}_{l,m}(x)
\hspace{20mm}A^{-}_{m}A^{+}_{m}B^{(q,\beta)}_{l,m-1}(x)={\cal E}_{l,m}\,B^{(q,\beta)}_{l,m-1}(x).
\end{eqnarray}
In this paper, we are interested to present the raising and
lowering relations of the indices $l$ and $m$ of the associated Bessel functions for both regions,
$l<0$ and $l\geq 0$ of $m-\frac{q}{2}\leq l\leq \frac{q}{2}-m-1$.
The shape invariance equations with respect to $l$ and $m$ are realized for every
normalization coefficient. However, realization of the laddering equations
\renewcommand\theequation{\arabic{equation}}
\setcounter{equation}{\value {tempeq}}
\setcounter{tempeq}{\value{equation}}
\renewcommand\theequation{\arabic{tempeq}\alph{equation}}
\setcounter{equation}{0} \addtocounter{tempeq}{1}
\begin{eqnarray}
&&\hspace{-10mm}A^{+}_{l,m}B^{(q,\,\beta)}_{l-1,m}(x)=\sqrt{E_{l,m}}\,B^{(q,\,\beta)}_{l,m}(x)
\hspace{15mm}A^{-}_{l,m}B^{(q,\,\beta)}_{l,m}(x)=\sqrt{E_{l,m}}\,B^{(q,\,\beta)}_{l-1,m}(x),\\
&&\hspace{-10mm}A^{+}_{m}B^{(q,\,\beta)}_{l,m-1}(x)=\sqrt{{\cal E}_{l,m}}\,B^{(q,\beta)}_{l,m}(x)
\hspace{18mm}A^{-}_{m}B^{(q,\beta)}_{l,m}(x)=\sqrt{{\cal E}_{l,m}}\,B^{(q,\beta)}_{l,m-1}(x),
\end{eqnarray}
imposes two recursion relations with respect to $l$ and $m$, respectively, on the
coefficients $a_{l,m}(q,\beta)$. One can easily show that the laddering equations (12a) and (12b)
as well as their corresponding recursion relations for the normalization coefficients
$a_{l,m}{(q,\beta)}$, are simultaneously established if the latter are chosen as
\begin{eqnarray}
\nonumber
\hspace{7mm}
a_{l,m}(q,\beta)=\left\{
\begin{array}{llllrrr}
\frac{\beta^{-l}\left(-1\right)^{\frac{q}{2}-m}}{\sqrt{\Gamma\left(l-m+\frac{q}{2}+1\right)\Gamma\left(-l-m+\frac{q}{2}\right)}}\hspace{20mm}m-\frac{q}{2}\leq l<0 &&&&\hspace{12mm}\mbox{(13a)} \\
\frac{\beta^{-\,l-1}\left(-1\right)^{-l-m+\frac{q}{2}-1}}{\sqrt{\Gamma\left(l-m+\frac{q}{2}+1\right)\Gamma\left(-l-m+\frac{q}{2}\right)}}\hspace{20mm}0\leq l\leq \frac{q}{2}-m-1,&&&&\hspace{12mm}\mbox{(13b)}
\end{array}\right.
\end{eqnarray}
where $m\leq \frac{q-1}{2}$.
One cannot actually follow the raising and lowering form from shape invariance symmetries
by means of the arbitrary normalization coefficients. The point is in order to realize
the laddering symmetries (12a) and (12b), it is a necessary  condition to select
the normalization coefficients as in (13a) and (13b).

Our reason for making the associated Bessel functions with $l<0$ goes back to the fact that
the functions $B^{(q,\,\beta)}_{0,m}(x)$ with $m\leq \frac{q-1}{2}$ are not annihilated by
the operators $A^{-}_{0,m}$.
Therefore, for a given $m$, decreasing of the index $l$ can be terminated at $l=m-\frac{q}{2}$.
Indeed, from the Eqs. (12a) we have $A^{+}_{\frac{q}{2}-m,m}B^{(q,\,\beta)}_{\frac{q}{2}-m-1,m}(x)=0$ and
$A^{-}_{m-\frac{q}{2},m}B^{(q,\,\beta)}_{m-\frac{q}{2},m}(x)=0$.
Moreover, from the Eqs. (12b), it becomes clear that the
associated Bessel functions lain on the lines  $l=m-\frac{q}{2}$  and $l=-m+\frac{q}{2}-1$
are annihilated by the ladder operators shifting $m$:
$A^{+}_{l+\frac{q}{2}+1}B^{(q,\,\beta)}_{l,l+\frac{q}{2}}(x)=0$
and $A^{+}_{\frac{q}{2}-l}B^{(q,\,\beta)}_{l,\frac{q}{2}-l-1}(x)=0$.
According to the above discussions, for a  given nonnegative $l$ and for
every $m\leq \frac{q-1}{2}$, we have $B^{(q,\,\beta)}_{-l-1,m}(x)=\beta(-1)^{l+1}B^{(q,\,\beta)}_{l,m}(x)$,
which means the functions with $l<0$ and $l\geq 0$ settled on the horizontal
lines of the Fig. 1 are mutually dependent on each other.
However, each of the vertical and oblique lines possess the associated Bessel
functions that are independent of each other.
This allows us to construct the generating functions for
them in three different methods by using
Rodrigues formula (4) for both regions $l<0$ and $l\geq 0$.
Note that the differential equation (5) has two independent solutions and above
discussions have focused on one of them.

\section{Exponential generating functions for the associated Bessel functions}
The square integrable associate Bessel functions can be applied to
obtain bound states corresponding to some one-dimensional supersymmetric potentials and also
some two-dimensional quantum mechanical models
having a Lie algebra symmetry \cite{Fakhri1,Fakhri3}.  Therefore, exponential generating
functions corresponding to the formal power series of
associate Bessel functions $B^{(q,\,\beta)}_{l,m}(x)$ with the same $l$, the same $l+m$ and the same $l-m$,
are important not only from the point of view of  mathematical derivation but also from the point of view of
physical applications. What we do is to consider
three different methods for computing the generating functions,
based on the presentation of appropriate infinite sequences of the associated Bessel functions.
The first type of the infinite sequences is $\{B^{(q,\,\beta)}_{l,m}(x)\}_{m=\frac{q}{2}+l}^{-\infty}$ for $l<0$
or $\{B^{(q,\,\beta)}_{l,m}(x)\}_{m=\frac{q}{2}-l-1}^{-\infty}$ for $l\geq0$.
Using the definitions $n:=\frac{q}{2}-l-m-1$ and $p:=l-m+\frac{q}{2}$,
all the associated Bessel functions $B^{(q,\,\beta)}_{l,m}(x)$ with $l<0$ and $l\geq0$
can also be rearranged as the union of the second and third type
infinite sequences: $\{B^{(q,\,\beta)}_{l,-l-n+\frac{q}{2}-1}(x)\}_{n=0}^{\infty}$ and
$\{B^{(q,\,\beta)}_{l,l-p+\frac{q}{2}}(x)\}_{p=0}^{\infty}$, respectively.
The second type sequences are increasing with respect to the index $l$ of
the associated Bessel functions $B^{(q,\,\beta)}_{l,m}(x)$ while
the third type sequences are decreasing.
These sequences are automatically
covered by the associated Bessel functions which
are linearly independent on the interval $0 \leq x<\infty$ with respect to the inner product (8), as
$\{B^{(q,\,\beta)}_{\frac{q}{2}-m-2k-2,m}(x)\}_{m=\frac{q}{2}-k-1}^{-\infty}\bigcup
\{B^{(q,\,\beta)}_{\frac{q}{2}-m-2k-1,m}(x)\}_{m=\frac{q}{2}-k-1}^{-\infty}$ and
$\{B^{(q,\,\beta)}_{m+2k-\frac{q}{2}+1,m}(x)\}_{m=\frac{q}{2}-k-1}^{-\infty}\bigcup
\{B^{(q,\,\beta)}_{m+2k-\frac{q}{2},m}(x)\}_{m=\frac{q}{2}-k-1}^{-\infty}$ with
$k=0,1,2,\cdots$.
Now we are in a position that for the sequences given in above we can derive
three different types of the generating functions.
\newline
\newline
{\bf Generating functions for given $q$ and $l$:}
Let us first introduce the generating functions for the power series with
different $m$ but with the same $l$, which are calculated  in a way similar to those for generalized Bessel
polynomials \cite{Srivastava3,Srivastava4,Lin}. The generating functions corresponding to the first
type sequences of associated Bessel functions  with the
different $m$ but the same $l$,
\renewcommand\theequation{\arabic{equation}}
\setcounter{equation}{\value {tempeq}}
\begin{eqnarray}\nonumber
\hspace{12.4mm}G_{l}(x,t):=\left\{\begin{array}{llllllll}
\sum_{m=0}^{\infty}\frac{t^{m}}{m!}
\frac{B^{(q,\,\beta)}_{l,l-m+\frac{q}{2}}(x)}{a_{l,l-m+\frac{q}{2}}(q,\beta)}
&&\hspace{30mm}\mbox{for}\hspace{3mm}l<0
\hspace{28mm}\mbox{(14a)}
\\
\sum_{m=1}^{\infty}\frac{t^{m}}{m!}
\frac{B^{(q,\,\beta)}_{l,l-m+\frac{q}{2}}(x)}{a_{l,l-m+\frac{q}{2}}(q,\beta)}
&&\hspace{30mm}\mbox{for}\hspace{3mm}l\geq0,
\hspace{26mm}\mbox{(14b)}
\end{array}
\right.
\end{eqnarray}
for $\left|t\right|<1$, are
\setcounter{equation}{14}
\begin{eqnarray}
\hspace{-6mm}G_{l}(x,t)=\left(1+t\right)^{2l}x^{l+\frac{q}{2}}e^{\frac{\beta
t}{x(1+t)}}+\lim_{z\rightarrow x}{\left[\frac{1}{\Gamma(2l+1)}\left(e^{\frac{\beta}{x}}\right)\left(\frac{d}{dz}\right)^{2l}\left(\frac{z^{2l}e^{\frac{-\beta}{x}}}{z-x(t+1)}\right)\right]}\nonumber\\
&&\hspace{-134mm}=\left(1+t\right)^{2l}x^{l+\frac{q}{2}}e^{\frac{\beta
t}{x(1+t)}}.
\end{eqnarray}
The relation (15), for both regions $l<0$ and $l\geq0$,  are followed by  the following
Cauchy's integral formula
\begin{eqnarray}
&&\hspace{-10mm}
\frac{1}{\Gamma(k+1)}\left(\frac{d}{dx}\right)^{k}\left(x^{2l}e^{\frac{-\beta}{x}}\right)=
\frac{1}{2\pi i}\oint_{{\cal C}(x,t)} dz \frac{z^{2l}e^{\frac{-\beta}{z}}}{(z-x)^{k+1}},
\end{eqnarray}
with $k=l-m+\frac{q}{2}$.
The Fig. 2 shows that the integration contour ${\cal C}(x,t)$ is a closed path around the circle
$\left|z-x\right|=\left|t\right|x$ in positive direction.
$z=0$ has also been settled out of the contour ${\cal C}(x,t)$.
Also, the real pole $z=x(1+t)$ is always settled inside the contour, since it is on the circle
at the right and left hand sides of center whether $t$ is positive or negative.
Considering the following relation
\begin{eqnarray}
B^{(q,\beta)}_{l,m}(x)=a_{l,m}(q,\beta)(-1)^{l-m+\frac{q}{2}}x^{l+\frac{q}{2}}\Gamma(l-m+\frac{q}{2}+1)
L_{l-m+\frac{q}{2}}^{-2l-1}\left(\frac{\beta}{x}\right),
\end{eqnarray}
one can establish the connection between the $G_{l}(x,t)$ with the Laguerre generating function (1.18)
of Ref. \cite{Lin}.
\newline
\newline
{\bf Generating functions for given $q$ and $l+m$:} In this case the sequences are increasing with
respect to the $l$. Due to the fact that whether $n$ is odd or even, i.e. $n=2k+1$ or $n=2k$,
highest functions are $B^{(q,\,\beta)}_{-k-1,\frac{q}{2}-k-1}(x)$ and
$B^{(q,\,\beta)}_{-k,\frac{q}{2}-k-1}(x)$, respectively.
These functions lie on the lines $m=l+\frac{q}{2}$ and $m=l+\frac{q}{2}-1$ of Fig. 1, respectively.
Therefore, it is obvious that the terminology of highest functions
has been devoted to the associated Bessel functions $B^{(q,\,\beta)}_{l,m}(x)$ with the most value for $m$.
{\bf a)} First we suppose that $n$ is odd, i.e. $n=2k+1$. For a given value of $k$,
the generating functions corresponding to the second type sequences are calculated as
\begin{eqnarray}
&&\hspace{-8mm}G_{n=2k+1}(x,t)=\sum_{m=0}^{\infty}\frac{t^{m}}{(2m)!}
\frac{B^{(q,\,\beta)}_{m-k-1,\frac{q}{2}-m-k-1}(x)}{a_{m-k-1,\frac{q}{2}-m-k-1}(q,\beta)}\nonumber\\
&&\hspace{17mm}=x^{\frac{q}{2}+k+1}e^{\frac{\beta}{x}}\sum_{m=0}^{\infty}\frac{(xt)^{m}}{(2m)!}
\left(\frac{d}{dx}\right)^{2m}\left(x^{2m-2k-2}e^{\frac{-\beta}{x}}\right)
\nonumber\\
&&\hspace{17mm}=x^{\frac{q}{2}+k+1}e^{\frac{\beta}{x}}\sum_{m=0}^{\infty}\frac{(xt)^{m}}{2\pi i}
\oint_{C(x,t)} dz \frac{z^{2m-2k-2}e^{\frac{-\beta}{z}}}{(z-x)^{2m+1}}
\nonumber\\
&&\hspace{17mm}=\frac{x^{\frac{q}{2}+k+1}e^{\frac{\beta}{x}}}{4\pi i\, x\sqrt{xt}}
\left[\oint_{C(x,t)} dz \frac{z^{-2k-2}(z-x)e^{\frac{-\beta}{z}}}{z-\frac{x}{1-\sqrt{xt}}}
-\oint_{C(x,t)} dz \frac{z^{-2k-2}(z-x)e^{\frac{-\beta}{z}}}{z-\frac{x}{1+\sqrt{xt}}}\right]
\nonumber\\
&&\hspace{17mm}=\frac{1}{2}x^{\frac{q}{2}-k-1}
\left[\left(1-\sqrt{xt}\right)^{2k+1}e^{\beta\sqrt{\frac{t}{x}}}
+\left(1+\sqrt{xt}\right)^{2k+1}e^{-\beta\sqrt{\frac{t}{x}}}\right].
\end{eqnarray}
As it has been shown in Fig. 3, $C(x,t)$ is a closed contour in positive direction
around a circle with center at $\left(\frac{x}{1-xt},0\right)$ and radius of $R=\frac{x\sqrt{xt}}{1-xt}$
on the complex plane $z$. Again, $z=0$ is out of the contour $C(x,t)$.
Also, the variable $t$ is considered as $t<\frac{1}{x}$. Therefore, $z=x$ settles on the real axis
at the inside of the circle and at the left hand side of the center. Moreover,
the poles $z=\frac{x}{1-\sqrt{xt}}$ and $z=\frac{x}{1+\sqrt{xt}}$ have been settled on the real axis
as well as on the circle,  at the right and left hand sides of center, respectively.
{\bf b)} Second we suppose that $n$ is even, i.e. $n=2k$. For a given value of $k$,
once again using the integration contour $C(x,t)$,
the generating functions of the second type sequences are calculated as
\begin{eqnarray}
&&\hspace{-8mm}G_{n=2k}(x,t)=\sum_{m=0}^{\infty}\frac{t^{m}}{(2m+1)!}
\frac{B^{(q,\,\beta)}_{m-k,\frac{q}{2}-m-k-1}(x)}{a_{m-k,\frac{q}{2}-m-k-1}(q,\beta)}\nonumber\\
&&\hspace{13.5mm}=\frac{x^{\frac{q}{2}-k}}{2\sqrt{xt}}
\left[\left(1-\sqrt{xt}\right)^{2k}e^{\beta\sqrt{\frac{t}{x}}}
-\left(1+\sqrt{xt}\right)^{2k}e^{-\beta\sqrt{\frac{t}{x}}}\right].
\end{eqnarray}
Note that the series in (18) and (19) are summed on the parameter $m$ for given values
$\frac{q}{2}-2k-2$ and $\frac{q}{2}-2k-1$ of $l+m$, respectively.
\newline\newline
{\bf Generating functions for given $q$ and $l-m$:}
This case involves  the decreasing sequences with
respect to the $l$ and again, there exists two possibilities:
$p$ can be both odd and even. For $p=2k+1$ and $p=2k$,
the highest functions corresponding to them are $B^{(q,\,\beta)}_{k,\frac{q}{2}-k-1}(x)$ and
$B^{(q,\,\beta)}_{k-1,\frac{q}{2}-k-1}(x)$, respectively.
Also, these functions have been placed
on the lines $m=-l+\frac{q}{2}-1$ and $m=-l+\frac{q}{2}-2$ of Fig. 1, respectively.
{\bf a)} In the case that $p$ is odd, i.e. $p=2k+1$, for a given value of $k$ and for $\left|t\right|<\infty$,
the generating functions of third type sequences are calculated as follows
\begin{eqnarray}
&&\hspace{-8mm}G_{p=2k+1}(x,t)=\sum_{m=0}^{\infty}\frac{t^{m}}{m!}
\frac{B^{(q,\,\beta)}_{k-m,\frac{q}{2}-m-k-1}(x)}{a_{k-m,\frac{q}{2}-m-k-1}(q,\beta)}\nonumber\\
&&\hspace{17mm}=(2k+1)!\, x^{\frac{q}{2}+k+1}e^{\frac{\beta}{x}}\sum_{m=0}^{\infty}\frac{(xt)^{m}}{2\pi i\, m!}
\oint_{C(x)} dz \frac{z^{2k-2m}e^{\frac{-\beta}{z}}}{(z-x)^{2k+2}}
\nonumber\\
&&\hspace{17mm}=x^{\frac{q}{2}+k+1}e^{\frac{\beta}{x}}\left[\left(\frac{d}{dz}\right)^{2k+1}
\left(z^{2k}e^{\frac{tx}{z^2}-\frac{\beta}{z}}\right)\right]_{z=x}.
\end{eqnarray}
In order to satisfy Eq. (20), it is sufficient that the arbitrary
contour $C(x)$  is chosen so that the points $z=x$ and $z=0$
lay inside and outside of that, respectively (see Fig. 4). {\bf b)}
When $p$ is an odd number, i.e. $p=2k$, the generating functions
of third type sequences for a given $k$ are calculated as follows
\begin{eqnarray}
&&\hspace{-8mm}G_{p=2k}(x,t)=\sum_{m=0}^{\infty}\frac{t^{m}}{m!}
\frac{B^{(q,\,\beta)}_{k-m-1,\frac{q}{2}-m-k-1}(x)}{a_{k-m-1,\frac{q}{2}-m-k-1}(q,\beta)}\nonumber\\
&&\hspace{13.5mm}=(2k)!\, x^{\frac{q}{2}+k+1}e^{\frac{\beta}{x}}\sum_{m=0}^{\infty}\frac{(xt)^{m}}{2\pi i\, m!}
\oint_{C(x)} dz \frac{z^{2k-2m-2}e^{\frac{-\beta}{z}}}{(z-x)^{2k+1}}
\nonumber\\
&&\hspace{13.5mm}=x^{\frac{q}{2}+k+1}e^{\frac{\beta}{x}}\left[\left(\frac{d}{dz}\right)^{2k}
\left(z^{2k-2}e^{\frac{tx}{z^2}-\frac{\beta}{z}}\right)\right]_{z=x}.
\end{eqnarray}
Here, the series in (20) and (21) are summed on the parameter $m$ for given values
$2k-\frac{q}{2}+1$ and $2k-\frac{q}{2}$ of $l-m$, respectively.
The accordance of the above generating functions with the theorem 1 of Ref. \cite{Srivastava4}
can be considered as new confirmation for it.

If we choose $q=0$, then we can claim that the relations (18), (19), (20) and (21)
are generating functions corresponding to the associated Bessel functions with
$l+m=-2(k+1)$, $l+m=-2k-1$, $l-m=2k+1$ and $l-m=2k$, respectively.
Therefore, we have obtained four new different types of generating
functions for the associated Bessel functions
depending on whether $l+m$ and $l-m$ are negative even or negative odd integers
and positive odd or nonnegative even integers, respectively.
Therefore, in order to obtain new generating functions
we have used square integrable associated Bessel functions in both regions $l<0$ and $l\geq 0$ with
the same Rodrigues representations for them.

\hspace{-8.2mm}
Fig. 1. The comprehensive plan of the squared integrable solutions for the differential equation (3) of
associated Bessel functions with $q=6$.
\newline\newline\newline\newline
Fig. 2. Plot of the integration contour ${\cal C}(x,t)$ for the generating functions of
first type sequences of the associated Bessel functions.
\newline\newline\newline\newline
Fig. 3. Plot of the integration contour $C(x,t)$ for the generating functions of
second type sequences of the associated Bessel functions.
\newline\newline\newline\newline
Fig. 4. Plot of the integration contour $C(x)$ for the generating functions of
third type sequences of the associated Bessel functions.

\end{document}